\documentclass[preprint,pteplogo]{ptephy_v2}

\usepackage{hyperref}

\usepackage{amsmath} 
\usepackage{graphicx}
\usepackage{subfigure} 
\usepackage{url} 

\begin{document}

\title{Forward neutrino production and event rates at the Future Circular Collider for hadron collisions}


\author[1]{E. Won}
\affil{Department of Physics, Korea University, 145 Anam-ro, Seoul,
  02841, Republic of Korea \email{eunilwon@korea.ac.kr}}
\author[2]{B. R. Ko\thanks{corresponding author}}
\affil{Department of Accelerator Science, Korea University Sejong
  Campus, 2511 Sejong-ro, Sejong, 30019, Republic of Korea \email{brko@korea.ac.kr}}




\begin{abstract}%
  A proposed future ultra high energy collider such as the Future
  Circular Collider is expected to produce intense and collimated
  neutrino beams from weak hadron decays. In this study, we estimate
  the production yields of such neutrinos from proton-proton
  collisions at a center-of-mass energy of 100~TeV and an integrated
  luminosity of 1~ab$^{-1}$.  
  Based on a hypothetical detector positioned either 0.5~km or 2~km
  downstream from the interaction point along the beamline, we
  derive the expected rates of charged current neutrino scattering
  events and their extension to neutrino energies of up to 50~TeV. We,
  for the first time, also evaluate the feasibility of observing the
  experimentally unverified direct production of $W^{\pm}$ bosons from
  neutrino-nucleus interactions.  
\end{abstract}

\subjectindex{C01, C32}

\maketitle

\section{Introduction}\label{introduction}
A large part of the particle physics community is dedicating growing
efforts towards the construction of higher-energy colliders that
exceed the capabilities of the Large Hadron Collider
(LHC)~\cite{LHC1,LHC2} to explore physics beyond the Standard Model
(SM), which has not yet been evidenced by the ATLAS~\cite{ATLASexp}
and CMS~\cite{CMSexp} experiments.
One such proposed facility is the Future Circular Collider (FCC)
project~\cite{FCC_V1}. The FCC is subdivided into three collider
types: FCC-hh for hadron-hadron collisions (with an option for
heavy-ion collisions), FCC-ee for electron-positron collisions, and
FCC-eh for electron-hadron collisions. Among these, the FCC-hh is
expected to operate  at a center-of-mass energy of $\mathcal{O}$(100)
TeV with a target average instantaneous luminosity of
$\mathcal{O}(10^{35})~\textrm{cm}^{-2} \textrm{s}^{-1}$
\cite{LUMI-FCC-hh}.

Such high-energy, high-luminosity hadron colliders are known to
produce intense neutrino and antineutrino beams via weak hadron
decays. At the LHC, neutrino fluxes and corresponding event rates have
been estimated~\cite{NuFluxLHC1, NuFluxLHC2}, and the relevant
experimental observations by the FASER
collaboration~\cite{FASER_observation, FASER_cross, FASER_muon} and
the SND@LHC
collaboration~\cite{SNDLHC_observation1,SNDLHC_observation2} have been
recently reported, where the former and the latter made their
observations on the beam collision axis and slightly off-axis,
respectively.
A recent study~\cite{FPFFCC} broadened these neutrino physics
observations by examining various detector configurations at a
100~TeV proton-proton ($pp$) collider and by outlining their potential
to probe new physics. 

In this study, we examine configurations involving a FASER$\nu$-like
detector with a mass of 128~kg used for ref.~\cite{FASER_cross}, a
subset of the entire FASER detector target whose mass is
1100~kg~\cite{FASER_cross}, positioned 0.5~km or 2~km downstream of
the $pp$ interaction point along the beamline of the FCC-hh.
In addition to evaluating the expected rates of charged current
neutrino scattering events for neutrino energies up to approximately
50~TeV, we explore the possibility of observing on-shell $W^\pm$ boson
production in neutrino-nucleus interactions, where hadronic coupling
is mediated by a virtual photon~\cite{ref:Wproduction1}.

In the FCC-hh forward region, a few tens of TeV energy neutrinos are
expected to be produced.
There have been no cross section measurements of neutrino-nucleus
interactions in this tremendous energy scale and such cross section
measurements allow us to test the SM prediction as always.
On top of that, this will be the unique place to observe on-shell
$W^\pm$ boson production in the collider environment, again an
excellent place to test the SM in another new way.

The $W^{\pm}$ boson production mechanism, mediated by a virtual
photon, allows both annihilation and exchange between incoming
neutrinos and interacting charged leptons.
Therefore, the $W^\pm$ production mechanism considered in this work is
complementary to $W^-$ production by the Glashow resonance mechanism
that permits the annihilation process only via
$\bar{\nu_e}~+~e^{-}\rightarrow~W^{-}$~\cite{Glashow} and, notably,
also permits $W^+$ production.
The Glashow resonance peaks sharply at 6.3~PeV using the nominal
masses of $W$ boson and electron, hence is practically insensitive at
the FCC-hh energy. On the other hand, the direct production of both
$W^-$ and $W^+$ from neutrino-nucleus scattering in this work can be
seen at the FCC-hh.
To date, a candidate Glashow resonance event has been reported by the
IceCube collaboration only~\cite{ICECUBE1, ICECUBE2}, hence it is
desirable to have follow up observations from different angles so that
can provide the sincere test of the SM.

As previously discussed in ref.~\cite{FPFFCC}, rigorous tests of the
SM involving neutrino interactions$-$including both deep inelastic
scattering and $W^{\pm}$ production, where the latter has never been
considered to date$-$may provide avenues for probing physics beyond
the SM (BSM), such as dark matter candidates, sterile neutrinos, and
non-standard neutrino interactions~\cite{ref:Machado}.

\section{Neutrino flux estimation}
Based on 1~ab$^{-1}$ of $\sqrt{s}=100$~TeV $pp$ collision simulation
data, as recorded by a detector at the FCC-hh, we estimated the
production yields of forward neutrinos using
\textsc{Pythia8}~\cite{PYTHIA6, PYTHIA8}, with parton distribution
functions (PDFs) obtained from LHAPDF~\cite{LHAPDF}.
\begin{table*}[h]
  \centering
  \begin{tabular}{l|l|lccc}\hline\hline
    &measurement&\textsc{Pythia8}&\texttt{14}&\texttt{19}&\texttt{20} \\ \hline
    $\sigma_{\rm inel}$ &$78.1\pm3.0$~\cite{ATL-inelastic}& \texttt{SoftQCD:inelastic}&78.05&78.05&78.05 \\ \hline
    $\sigma_{c\bar{c}X}$&$13.43\pm1.07$~\cite{LHCb-ccbar} & \texttt{SoftQCD:inelastic} by the counting&14.24&13.54&14.58\\ 
                        & & \texttt{HardQCD:hardccbar}&5.34&9.53&12.25\\ \hline
    $\sigma_{b\bar{b}X}$&$0.60\pm0.06$~\cite{LHCb-bbbar} & \texttt{SoftQCD:inelastic} by the counting&0.83&0.79&0.99\\ 
                        & & \texttt{HardQCD:hardbbbar}&0.35&0.60&0.53\\ \hline\hline

  \end{tabular}
  \caption{Measured and predicted cross sections at $\sqrt{s}=13$~TeV,
    expressed in mb,
    where the error in each measurement is the total error.
    The values \texttt{14}, \texttt{19}, and \texttt{20} refer to 
    the \texttt{Tune:pp} options provided by \textsc{Pythia8}.}
  \label{TAB:XSECS}
\end{table*}
\begin{figure*}[h]
  \centering
  \includegraphics[width=0.95\textwidth]{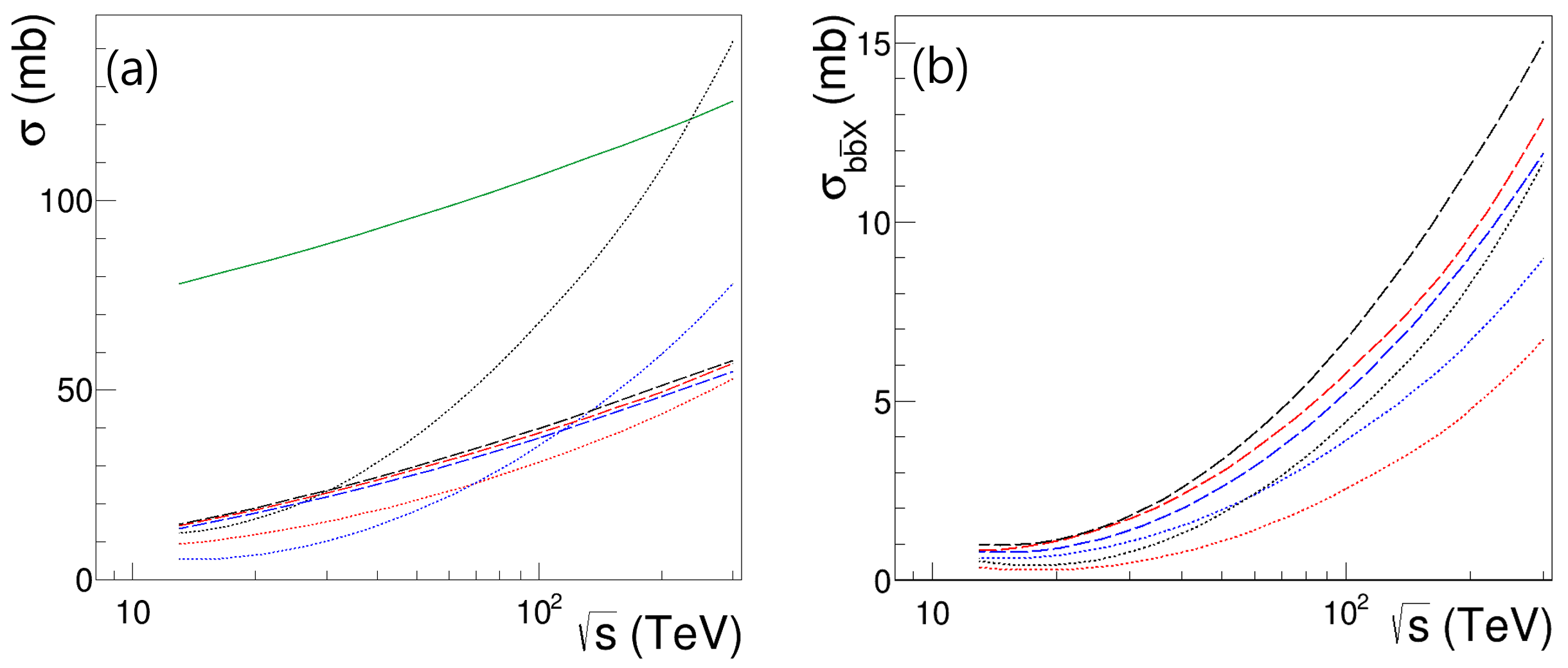}
  \caption{Panels (a) and (b) illustrate $\sigma_{c\bar{c}X}$ and
    $\sigma_{b\bar{b}X}$ as functions of the $pp$ collision energy
    $\sqrt{s}$. The green solid line in panel (a) represents
    $\sigma_{\rm inel}$ from \texttt{SoftQCD:inelastic}.
    The dashed lines correspond to cross sections derived from hadron
    counting, while the dotted lines in panels (a) and (b) represent
    the values obtained from \texttt{HardQCD:hardccbar} and
    \texttt{HardQCD:hardbbbar}, respectively. The red, blue, and black
    curves denote the results for \texttt{Tune:pp=14}, \texttt{19},
    and \texttt{20}, respectively.}  
  \label{FIG:XSECS}
\end{figure*}
The \texttt{SoftQCD:inelastic} process was employed, as its cross
section is largely insensitive to \textsc{Pythia8} parameter tuning
and demonstrates good agreement with the measured inelastic $pp$ cross
section
$\sigma_{\rm inel}=78.1\pm0.6\pm1.3\pm2.6$~mb~\cite{ATL-inelastic} at
$\sqrt{s}=13$~TeV.
Since \texttt{SoftQCD:inelastic} does not provide explicit cross
sections, we estimated the $\sigma_{b\bar{b}X}$ and
$\sigma_{{\rm prompt}~c\bar{c}X}$\footnote{We omit the term ``prompt''
when referring to prompt $c\bar{c}X$ events in the following
discussion.} by counting the numbers of charmed and $B$ hadrons in the
generated inelastic events.
To enable clearer identification of $c\bar{c}X$ events, $B$ hadron
decays were disabled during the estimation.
The resulting cross section estimates were then compared with those
obtained from \texttt{HardQCD:hardbbbar} and
\texttt{HardQCD:hardccbar}.
We performed the counting procedure using three \textsc{Pythia8} tunes:
\texttt{Tune:pp=14}~\cite{MONASH}, \texttt{19}~\cite{ATL14a}, and
\texttt{20}. \texttt{Tune:pp=14} corresponds to the default
configuration in \textsc{Pythia8}, whereas \texttt{Tune:pp=19} and
\texttt{20} are derived from the ``ATLAS A14 central
tune''~\cite{ATL14a} and use \texttt{CTEQ6L1} and \texttt{MSTW2008LO}
from LHAPDF~\cite{LHAPDF}, respectively.
\begin{figure*}[h]
  \centering
  \subfigure{\includegraphics[width=0.49\textwidth]{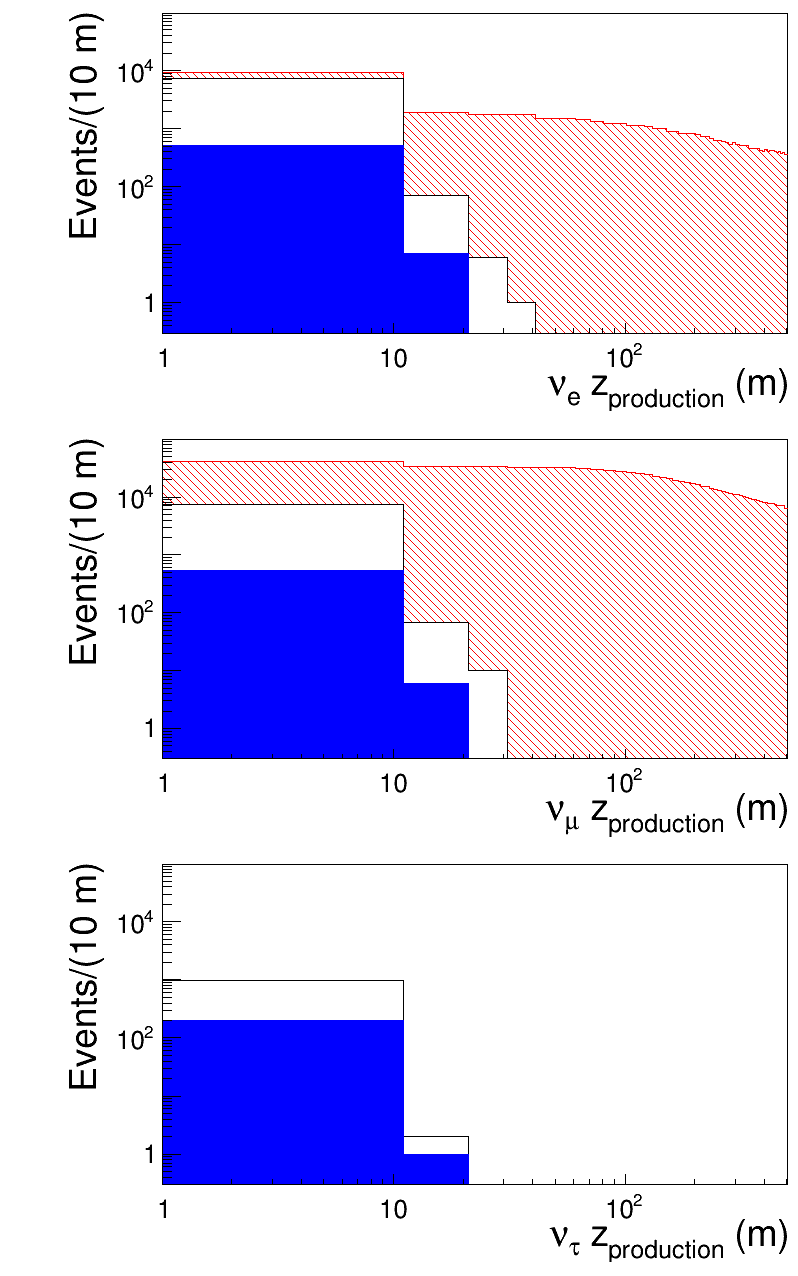}}
  \subfigure{\includegraphics[width=0.49\textwidth]{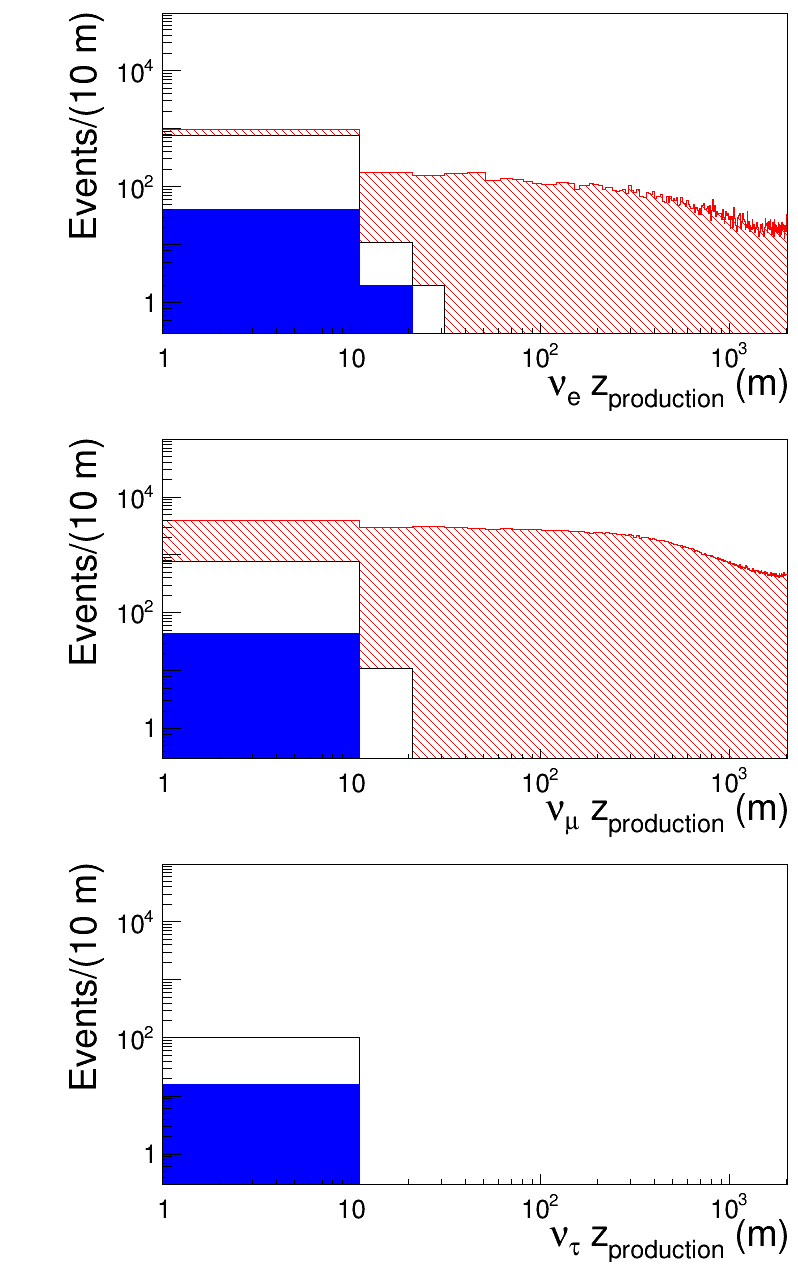}}
  \caption{Longitudinal production positions $|z_{\rm production}|$
    for $\nu_e$, $\nu_\mu$, and $\nu_\tau$, respectively, from top to
    bottom.
    Left and right panels show those when the target is positioned
    0.5~km and 2~km from the $pp$ collision point, respectively, where
    the target surface area is $23.4\times9$~cm$^2$.    
    The blue shaded, hollow, and red hatched are contributed from $B$,
    charmed, and light particles, respectively.
    Note that the events with $|z_{\rm production}|$ less than 1~m are
    not shown in the plots and the total event numbers are actually
    more than those shown in figure~\ref{FIG:Energy} later with the
    neutrino energy threshold of 0.5~TeV.}  
  \label{FIG:DZ0}
\end{figure*}
To improve agreement with measured cross
sections~\cite{LHCb-ccbar, LHCb-bbbar}, the charm quark mass $m_c$ was
adjusted from 1.5~GeV/$c^2$ to 1.0~GeV/$c^2$. Table~\ref{TAB:XSECS}
presents the predicted cross sections for each tune with modified
$m_c=$ 1.0~GeV/$c^2$, along with the corresponding experimental data.
The explicit cross sections from \texttt{HardQCD:hardbbbar} and
\texttt{HardQCD:hardccbar} are also included in
table~\ref{TAB:XSECS}. To extract the relevant cross section
$\sigma_{c\bar{c}X}$, we first reproduced the published multiplicative
factor of 3.9~\cite{LHCb-bbbar} using \textsc{Pythia8}, with this
factor accounting for the extrapolation of $\sigma_{b\bar{b}X}$ to the
full phase space. Adopting the same approach, we estimated the
corresponding multiplicative factor for $\sigma_{c\bar{c}X}$ in the
full phase space to be 4.73, yielding a full-phase-space cross section
of 13.43 mb from the measured value of
$2840\pm3\pm170\pm150$~$\mu$b~\cite{LHCb-ccbar}.
\begin{figure*}[h]
  \centering
  \subfigure{\includegraphics[width=0.49\textwidth]{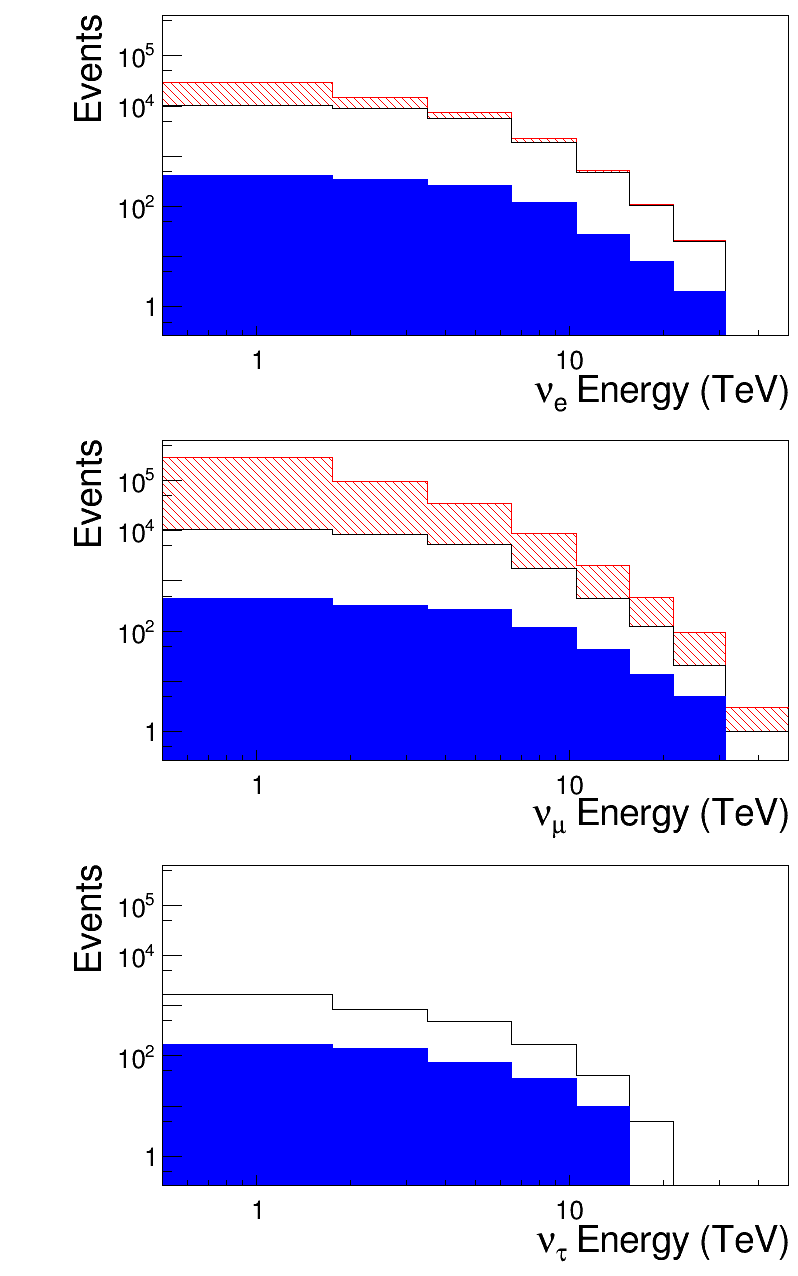}}
  \subfigure{\includegraphics[width=0.485\textwidth]{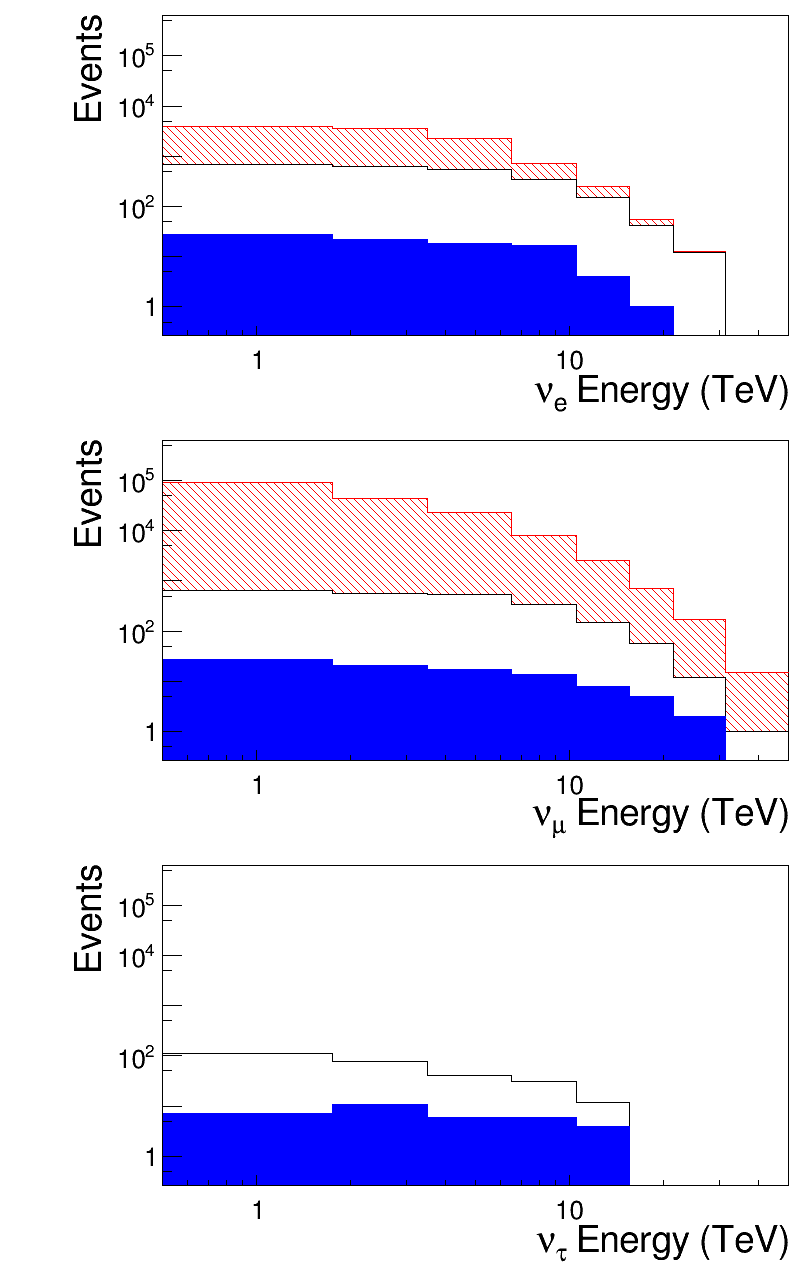}}
  \caption{Neutrino energy distributions above 0.5~TeV for $\nu_e$,
    $\nu_\mu$, and $\nu_\tau$, respectively, from top to bottom.
    Left and right panels show those when the target is positioned 0.5~km
    and 2~km from the $pp$ interaction point, respectively, where the
    target surface area is $23.4\times9$~cm$^2$.
    The blue shaded, hollow, and red hatched are contributed from $B$,
    charmed, and light particles, respectively.}
  \label{FIG:Energy}
\end{figure*}

According to the results in table~\ref{TAB:XSECS}, 
achieving simultaneous  agreement between the predicted
$\sigma_{c\bar{c}X}$ and $\sigma_{b\bar{b}X}$ values and experimental
measurements is non-trivial.
To identify an optimal and valid tuning configuration
at the FCC-hh collision energy, we
extended the predictions in table~\ref{TAB:XSECS} to $pp$
collision energies up to 300 TeV, as illustrated in figure~\ref{FIG:XSECS}.
The $\sigma_{c\bar{c}X}$ values from \texttt{HardQCD:hardccbar} exhibit 
a strong dependence on tuning parameters as $\sqrt{s}$ increases, with
one configuration yielding an unreasonable $\sigma_{c\bar{c}X}$ that
exceeds $\sigma_{\rm inel}$.
In contrast, the values obtained from \texttt{SoftQCD:inelastic} using
our counting method demonstrate reasonable scalability up to
$\sqrt{s}=300$~TeV, as depicted in figure~\ref{FIG:XSECS} (a).
Meanwhile, the $\sigma_{b\bar{b}X}$ values remain relatively stable 
across tuning options in \textsc{Pythia8} and do not show notable
deviations, as displayed in 
figure~\ref{FIG:XSECS} (b).
Based on the observed cross section trends in 
figure~\ref{FIG:XSECS}, we initially selected \texttt{Tune:pp=19} and
$m_c=1$~GeV/$c^2$. We then applied the \textsc{Pythia8} forward
physics tuning~\cite{ForwardTune} in combination with this configuration to
simulate high-energy neutrino produced along the beamline. 
Accordingly, all data used
for neutrino flux estimation at the FCC-hh
energy were generated using \texttt{Tune:pp=19},
$m_c=1$~GeV/$c^2$, and forward physics tuning.
The $\sigma_{\rm inel}$ value of 106.6~mb at $\sqrt{s}=100$~TeV, obtained 
from \texttt{SoftQCD:inelastic}, corresponds to the green solid line in
figure~\ref{FIG:XSECS} (a) and was adopted in this study.
Note that although we employed the $\sigma_{c\bar{c}X}$ and
$\sigma_{b\bar{b}X}$ values for our tuning, they are not dominant in
$\sigma_{\rm inel}$ in the end according to either measurements and
prediction (see table~\ref{TAB:XSECS} and figure~\ref{FIG:XSECS}),
thus not expecting significant contributions to our results in this
work.
  
For the neutrino flux estimation, $10^8$ inelastic $pp$ events
were generated at $\sqrt{s}=100$~TeV, and all resulting unstable
particles$-$including muons, pions, kaons, and neutrons$-$were allowed
to decay according to their lifetimes. Neutrinos were required to be
produced within 2.35~cm in the transverse direction, ensuring their
production occurred within the ATLAS beam pipe~\cite{ATL_IP}, in line
with the FASER experiment.
In this study, we applied the same surface area of the target ($S_T$)
used in the FASER experiment, $23.4\times9$~cm$^2$~\cite{FASER_cross},
when the detector target is positioned either 0.5~km or 2~km from the
$pp$ interaction point, respectively.
Figure~\ref{FIG:DZ0} illustrates the neutrino production position in
both the forward and backward longitudinal directions depending on the
positions of the detector target.
According to \textsc{Pythia8}, neutrinos produced from heavy-flavor
particles, such as charmed and $B$ hadrons, emerge within several tens
of meters, whereas those from light particles, including pions and
kaons, can originate at distances beyond 2~km, as can be conceived in
the right panel of figure~\ref{FIG:DZ0}.
In this simulation, charmed and kaon particles
primarily contribute to the $\nu_e$ flux, while the $\nu_\mu$ flux is
dominated by charmed, kaon, and pion particles.
\begin{table}[h]
  \centering
  \begin{tabular}{l|c|c}\hline\hline
    & $S_T=23.4\times9$~cm$^2$ & $S_T=23.4\times9$~cm$^2$\\ 
    & $z_{T}=0.5$~km &$z_{T}=2$~km\\
    & $0<z_{\rm production}<0.5$~km &$0<z_{\rm production}<2$~km\\ \hline
    $\nu_e$   &$2.92\times10^{13}$  &$5.71\times10^{12}$  \\ 
    $\nu_\mu$ &$2.30\times10^{14}$  &$9.06\times10^{13}$  \\ 
    $\nu_\tau$&$1.70\times10^{12}$  &$1.46\times10^{11}$\\ \hline\hline
  \end{tabular}
  \caption{Estimated numbers of neutrinos under different constraints
    at the FCC-hh, assuming equal production rates in the forward and
    backward regions and an integrated luminosity of 1~ab$^{-1}$ over
    one year. $z_{T}$ is the detector target location downstream from
    the $pp$ interaction point and $S_T$ is the surface area of the
    target.}  
  \label{TAB:NUFlux}
\end{table}

We calculated the transverse location of the
neutrino at the detector target location using the neutrino production
position and momentum information provided by
\textsc{Pythia8}. Neutrinos that hit the surface area of
$S_{T}=23.4\times9$~cm$^2$ at the detector location only are counted
as the neutrino fluxes.
Assuming that a FASER$\nu$-like detector is positioned 0.5~km or 2~km
downstream of the $pp$ collision point, figure~\ref{FIG:Energy} depicts 
the incoming neutrino energy spectra, which extend up to approximately
50~TeV.
Based on an integrated luminosity of 1~ab$^{-1}$ over one year at the
FCC-hh~\cite{LUMI-FCC-hh}, the resulting neutrino fluxes incident on the
detector target are listed inclusively in table~\ref{TAB:NUFlux} for
neutrinos with energies above 0.5~TeV.
\section{Neutrino event rates}
To estimate charged current neutrino interactions in the forward
region, we considered a FASER$\nu$-like detector equipped with
tungsten targets.
The detector had a cross-sectional area of
$23.4\times9$~cm$^2$ and a length of 31.6~cm, corresponding to a total
target mass of about 128~kg.
This configuration resulted in a target nucleon surface density of
approximately $N_T=3.65\times10^{26}$~nucleons/cm$^2$ and may have
enabled the use of the detection efficiencies employed by
FASER~\cite{FASER_cross}. The charged current neutrino scattering
cross section per nucleon, $\sigma$, was evaluated using~\cite{FERBEL}
\begin{eqnarray}
\sigma = \frac{\mathcal{N}}{\Phi N_T \epsilon},
\end{eqnarray}
where $\sigma=\sigma_{\nu}+\sigma_{\bar{\nu}}$ denotes the sum of the
neutrino and antineutrino cross sections per nucleon,
$\mathcal{N}$ denotes the neutrino event rate, $\Phi$ is the 
incident neutrino flux for the assumed integrated luminosity (e.g.,
the values in table~\ref{TAB:NUFlux}), and $\epsilon$ is the overall
detection efficiency.

\begin{figure*}[htbp]
  \centering
  \subfigure{\includegraphics[width=0.9\textwidth,height=0.85\textwidth]{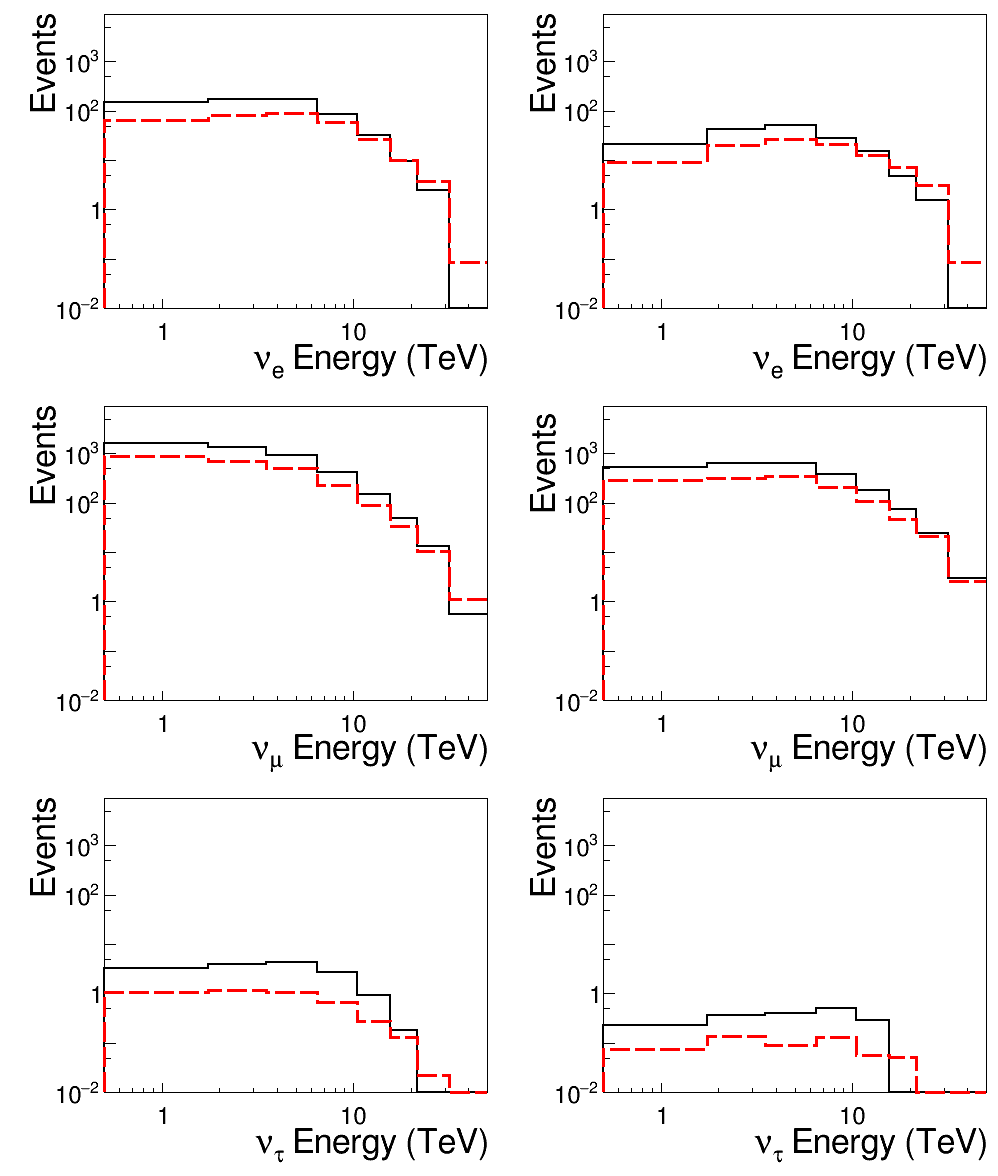}}
  \caption[xxx]{Expected numbers of the charged current interactions of
    scattered neutrino energy for $\nu_e$, $\nu_\mu$, and $\nu_\tau$,
    respectively, from top to bottom, where the black solid lines and
    red dashed lines are them from \textsc{Pythia8} and
    \texttt{EPOS.LHC-R}~\cite{EPOS.LHC-R},\footnotemark~respectively.
    The left and right panels are the cases with the detector
    locations of 0.5~km and 2~km, respectively.
    A subset of the entire FASER$\nu$ detector target corresponding to
    a mass of approximately 128~kg only was used for them.
  }  
  \label{FIG:N_cc}
\end{figure*}

The neutrino event rates $\mathcal{N}$ were estimated as a function
of neutrino energy $E$ in the range 0.5$-$50~TeV, using the
established relation 
$\sigma/E=1.0\times10^{-38}$~cm${^2}$GeV$^{-1}$~\cite{PDG}.
The hadronic showers from neutrino interactions may exceed the
containment capability of a FASER$\nu$-like detector at the FCC-hh
energy resulting in degrading the lepton identification efficiencies.
To account for that, we adopted a half of the FASER
efficiencies~\cite{FASER_cross}, thus $\epsilon=0.125$ for $\nu_e$ and
0.15 for $\nu_\mu$. 
In the absence of dedicated efficiency values for $\nu_{\tau}$
interactions, a lower efficiency $\epsilon=0.05$ was assumed for
$\nu_{\tau}$ events.
To examine the dependence of event rates $\mathcal{N}$ on
detector placement, two locations$-$0.5~km and 2~km from the FCC-hh
$pp$ interaction point$-$were considered.
Figure~\ref{FIG:N_cc} displays the charged current neutrino event rates
for $\nu_e$, $\nu_\mu$, and $\nu_\tau$ as functions of the scattered
neutrino energy, from top to bottom. The left and right panels
correspond to them when the detector is located at 0.5~km and 2~km,
respectively.
As illustrated in figure~\ref{FIG:N_cc} and consistent with the
incoming neutrino fluxes shown in figure~\ref{FIG:Energy}, the charged
current $\nu_\mu$ event rates are expected to reach energies
approaching 50~TeV.
Although higher energy neutrinos tend to be produced farther from the
interaction point, the corresponding event rates are reduced as
illustrated in figures~\ref{FIG:Energy} and \ref{FIG:N_cc}.
This is because as the detector locations get farther, the neutrino
beams have to get more collimated along the beamline to hit the
detector target whose surface area is generally constant.
\addtocounter{footnote}{-1}
\footnotetext[\thefootnote]{See section~\ref{VALID} for \texttt{EPOS.LHC-R}.}
\addtocounter{footnote}{-2}
\begin{figure*}[htbp]
  \centering
  \subfigure{\includegraphics[width=0.9\textwidth,height=0.85\textwidth]{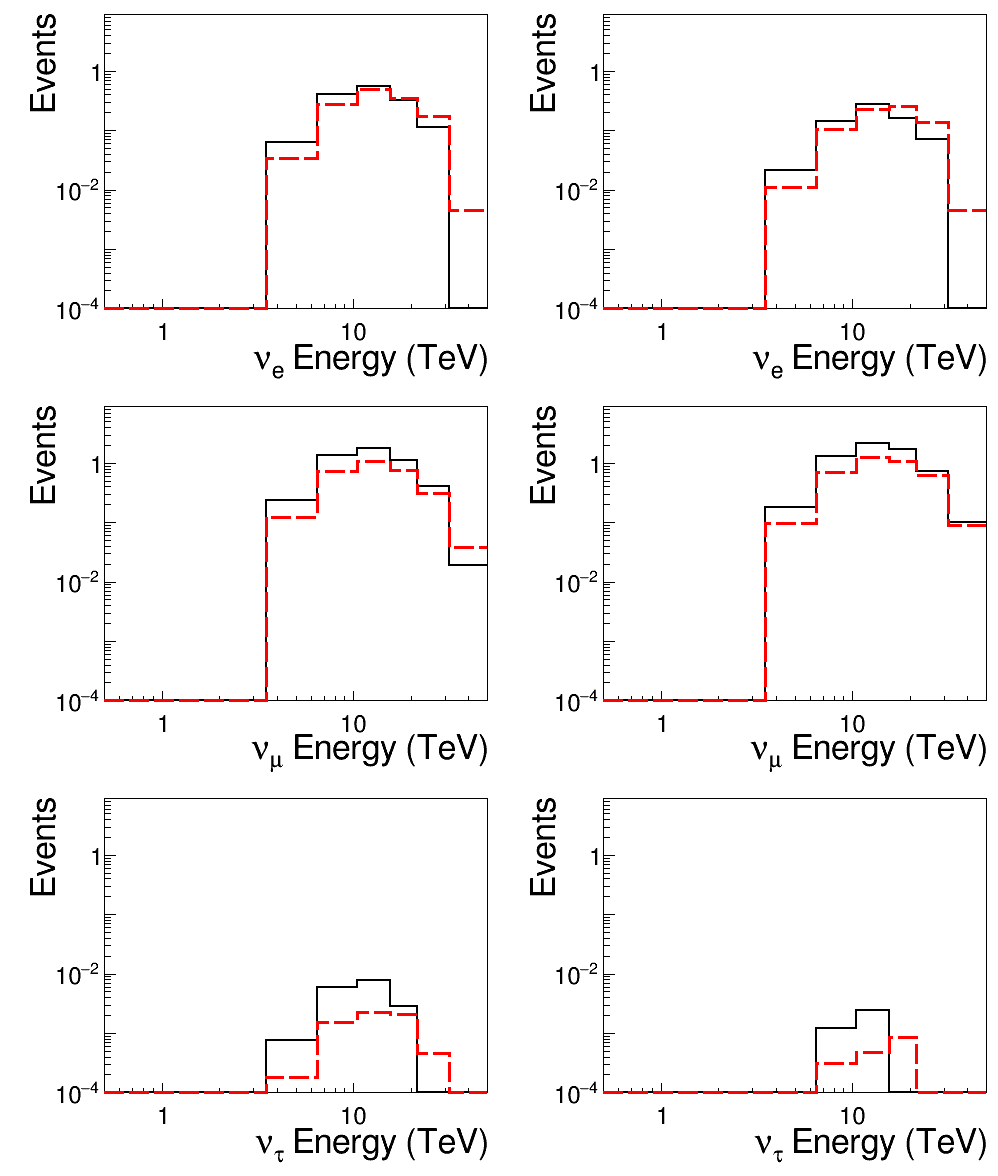}}
  \caption{Expected $W^{\pm}$ production of scattered neutrino energy
    for $\nu_e$, $\nu_\mu$, and $\nu_\tau$, respectively, from top to
    bottom, where the black solid lines and red dashed lines are them
    from \textsc{Pythia8} and \texttt{EPOS.LHC-R}, respectively.    
    The left and right panels are the cases with the detector
    locations of 0.5~km and 2~km, respectively.
    A subset of the entire FASER$\nu$ detector target corresponding to
    a mass of approximately 128~kg only was used for them.
  }  
\label{FIG:N_W}
\end{figure*}
\subsection{Direct production of $W^{\pm}$ bosons}
Using the same experimental parameters applied to estimate the neutrino
event rates in figure~\ref{FIG:N_cc}, we also evaluated 
direct $W^{\pm}$ boson production from neutrino-nucleus interactions,
where hadronic coupling is mediated by a virtual
photon~\cite{ref:Wproduction1}.
In this estimation, we employed the $W$ boson cross sections reported
in Fig.~12 of ref.~\cite{ref:Wproduction1} and assumed detection
efficiencies of a half those used to produce figure~\ref{FIG:N_cc} to
take account of the reconstruction of additional high energy charged
lepton, thus $\epsilon=6.25\%$, $7.5\%$, and $2.5\%$ for $\nu_e$,
$\nu_\mu$, and $\nu_\tau$, respectively.
The charged current neutrino scattering events shown in
figure~\ref{FIG:N_cc} are reconstructed from at least a single high
energy charged lepton identified in each event as done in
ref.~\cite{FASER_cross}.
Therefore, imposing the additional high energy charged lepton in each
event would improve the significance of $W$ boson events by isolating
them from the charged current neutrino scattering events.
In the end, the expected full-statistics dataset at the FCC-hh
(20$\--$30~ab$^{-1}$) would enable such separation between $W$ boson
events and the charged current neutrino scattering events even with
the lower detection efficiencies resulting from detector itself or
tighter event selection cuts.

As illustrated in figure~\ref{FIG:N_W}, several direct $W^{\pm}$
production is anticipated from $\nu_\mu$ interactions at the FCC-hh in
terms of the assumed FASER$\nu$-like detector geometry, 128~kg target
mass and $S_T=23.4\times9$~cm$^2$, and the assumed lower detection
efficiencies employed in this work.
Note that the $W$ boson production rates shown in figure~\ref{FIG:N_W}
are rather uniform over the relevant neutrino energy range, which
resulted from the $W$ boson cross sections reported in Fig.~12 of
ref.~\cite{ref:Wproduction1} that steadily increase as the neutrino
energies increase up to about 100~TeV, while the neutrino fluxes
decrease as the neutrino energies increase (see figure~\ref{FIG:N_cc}).
Thus, very high energy neutrinos at the FCC-hh can significantly
contribute to the direct $W$ production albeit their lower fluxes.
\subsection{Extensions to other detectors}
\begin{figure*}[h]
  \centering
  \subfigure{\includegraphics[width=0.9\textwidth,height=0.85\textwidth]{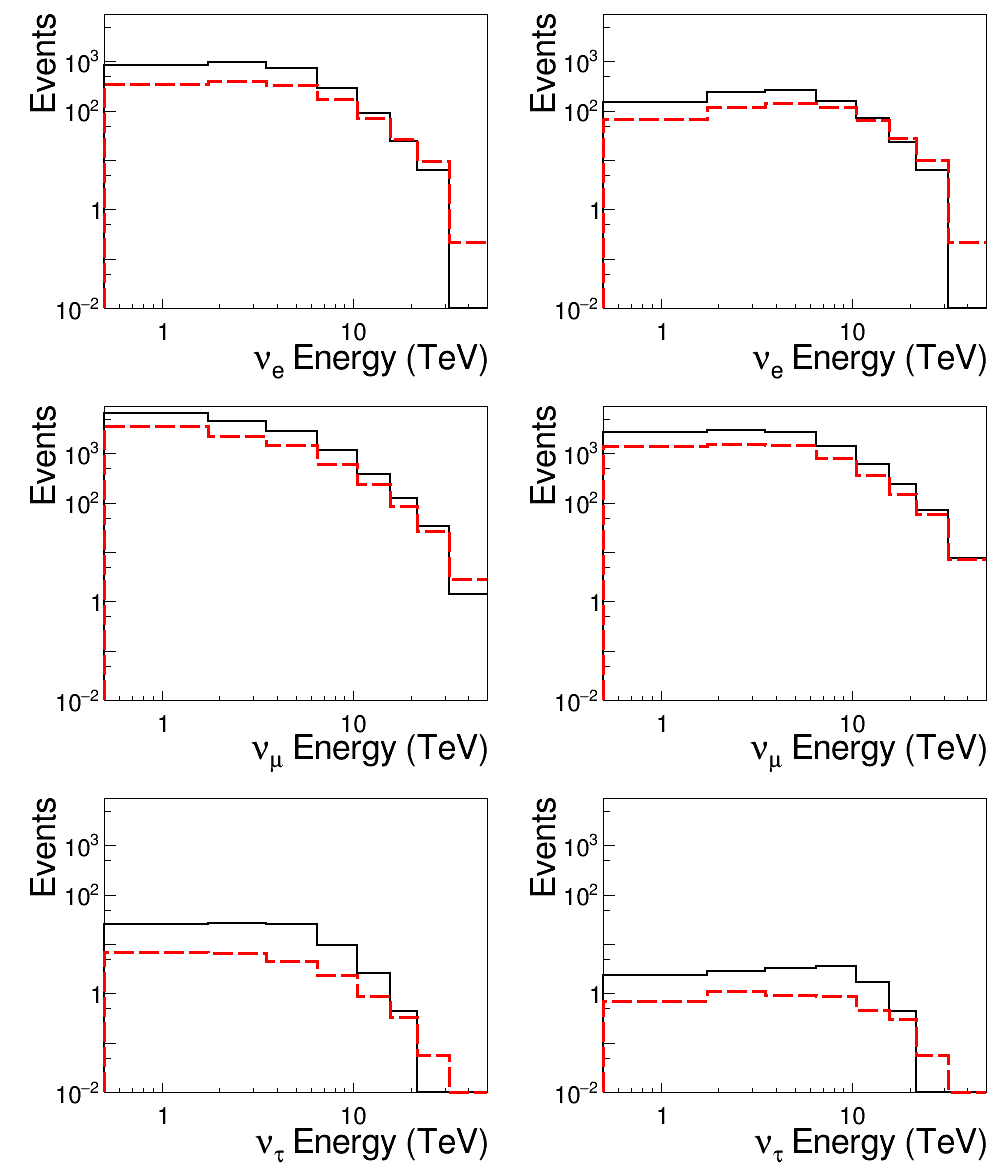}}
  \caption{The same as figure~\ref{FIG:N_cc} with the same detection
    efficiencies, but using the currently existing entire FASER$\nu$
    target whose specifications are a mass of 1100~kg,
    $S_T=25\times30$~cm$^2$, and a length of 80~cm resulting in an
    $N_T\simeq9.25\times10^{26}$~nucleons/cm$^2$.}  
  \label{FIG:N_cc_1100}
\end{figure*}
\begin{figure*}[h]
  \centering
  \subfigure{\includegraphics[width=0.9\textwidth,height=0.85\textwidth]{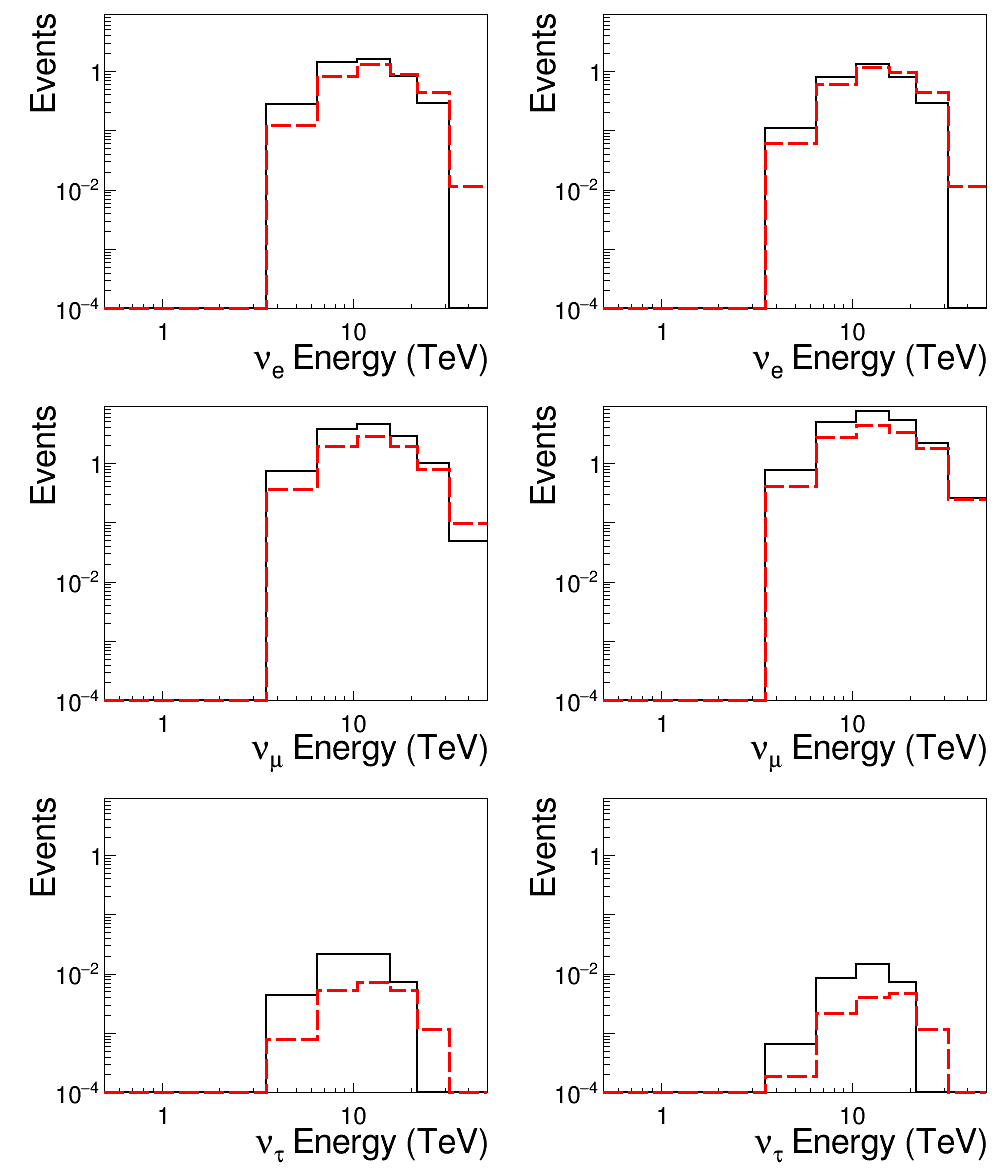}}
  \caption{The same as figure~\ref{FIG:N_W} with the same detection
    efficiencies, but using the currently existing entire FASER$\nu$
    target whose specifications are a mass of 1100~kg,
    $S_T=25\times30$~cm$^2$, and a length of 80~cm resulting in an
    $N_T\simeq9.25\times10^{26}$~nucleons/cm$^2$.}  
\label{FIG:N_W_1100}
\end{figure*}
The approach outlined so far can be straightforwardly extended to
other detectors with different target masses or alternative
placements, through appropriate adjustments to the target geometry.
Figures~\ref{FIG:N_cc_1100} and~\ref{FIG:N_W_1100} show the same as
figures~\ref{FIG:N_cc} and~\ref{FIG:N_W} assuming the same detection
efficiencies, respectively, but using the currently existing entire
FASER$\nu$ target whose specifications are a mass of 1100~kg,
$S_T=25\times30$~cm$^2$, and a length of 80 cm resulting in an
$N_T\simeq9.25\times10^{26}$~nucleons/cm$^2$.
As illustrated in figure~\ref{FIG:N_W_1100}, substantial direct $W^{\pm}$
production is anticipated from $\nu_\mu$ interactions at the
FCC-hh if the entire FASER$\nu$ detector target would be adopted.

A candidate Glashow resonance event, with a reported energy of
approximately 6.3~PeV$-$notably exceeding the FCC-hh energy$-$has been
observed by the IceCube collaboration~\cite{ICECUBE1, ICECUBE2}.
In spite of such an energy threshold at the FCC-hh, our results
herein, however, suggest that the FCC-hh may enable on-shell $W^{-}$
production, providing a complementary mechanism to the Glashow
resonance whose energy threshold lies well beyond the FCC-hh's
capability. Moreover, $W^{+}$ production is also expected at the
FCC-hh, in contrast to the Glashow resonance, which exclusively
produces $W^-$ bosons.

\section{Validations}\label{VALID}
Our estimations with \textsc{Pythia8} were validated by comparing them
not only with other predictions, but also with currently available
experimental data. Note that our validations herein are to make sure
of any strange behaviors in our estimations, like singularity, rather
than to assure the accuracy, because it is very difficult to discuss
accuracy in any estimations in the absence of experimental data. Note
also that this is the first comparison between different generators at
the expected FCC-hh center-of-mass energy, 100~TeV.

For other predictions, we used \texttt{EPOS.LHC-R}~\cite{EPOS.LHC-R}
with hadronic rescattering~\cite{RESCATTERING1, RESCATTERING2}
implemented in the \texttt{CRMC} simulation package~\cite{CRMC}, where
\texttt{EPOS.LHC-R} is the updated \texttt{EPOS.LHC}~\cite{EPOS.LHC}
to solve the muon puzzle~\cite{MUON1, MUON2}. The \texttt{EPOS.LHC-R}
results are already shown in
figures~\ref{FIG:N_cc},~\ref{FIG:N_W},~\ref{FIG:N_cc_1100},
and~\ref{FIG:N_W_1100} as the red dashed-lines.
The absolute event rates predicted from \textsc{Pythia8} and
\texttt{EPOS.LHC-R} are substantially different from each other as
shown in figures~\ref{FIG:N_cc},~\ref{FIG:N_W},~\ref{FIG:N_cc_1100},
and~\ref{FIG:N_W_1100}, while their behaviors are not too different
from each other.
Although the overall \texttt{EPOS.LHC-R} expectations are less than
those from \textsc{Pythia8}, the results with the entire FASER$\nu$
detector target running at the FCC-hh energy indicate that charged
current neutrino interaction rates can be expected for neutrino
energies reaching up to approximately 50~TeV and support a chance to
observe direct $W^-$ and $W^+$ boson production via neutrino-nucleus
interactions.

\begin{figure}[h]
  \centering
  \includegraphics[width=0.95\textwidth]{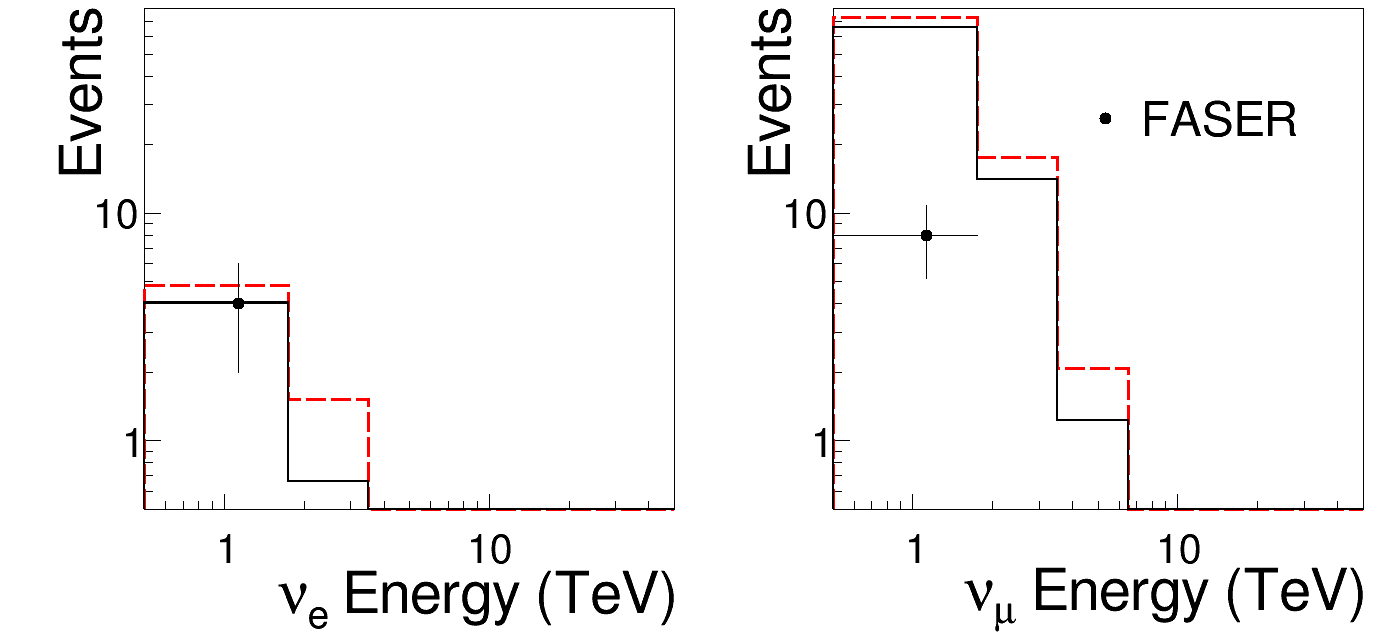}
  \caption{Expected charged interactions of scattered neutrino energy
    for $\nu_e$ (left) and $\nu_\mu$ (right), respectively, with our
    assumed FASER geometry, where the black solid lines and red dashed
    lines are them from \textsc{Pythia8} and \texttt{EPOS.LHC-R},
    respectively. The black solid circles with error bars are the
    observed numbers of neutrino events from the FASER$\nu$ detector.
    A subset of the entire FASER$\nu$ detector target corresponding to
    a mass of approximately 128~kg only was used for them.
  }  
  \label{FIG:FASER_byus}
\end{figure}

In order to validate our estimations further, we repeated the
procedure using the FASER parameters to reproduce the neutrino event
rates reported by the FASER collaboration~\cite{FASER_cross}. The
parameters adopted were  $\sqrt{s}=13.6$~TeV, an integrated luminosity
of 9.5~fb$^{-1}$, and an inelastic cross section
$\sigma_{\rm inel}=78.58$~mb at $\sqrt{s}=13.6$~TeV, which corresponds
to the green solid line in figure~\ref{FIG:XSECS} (a). We also applied
the FASER neutrino energy thresholds of 0.56 and 0.52~TeV for $\nu_e$
and $\nu_\mu$, respectively.
As depicted in figure~\ref{FIG:FASER_byus}, our predicted event
numbers are approximately 4 for $\nu_e$ and 60 for $\nu_\mu$ in the
relevant energy range, where the former is consistent with the FASER
measurement while the latter does not seem so.
For dissecting the comparison, we estimated the FASER numbers
corresponding to our numbers using the FASER simulation numbers in
Table~III in ref.~\cite{FASER_cross} and our simplified detection
efficiencies used in this work, and they are found to be 2 and 13 for
$\nu_e$ and $\nu_\mu$, respectively.
Therefore, our simulation results are double in the $\nu_\mu$ to
$\nu_e$ yield ratio and double in the $\nu_e$ yield compared with the
FASER simulation results, thus not substantially overestimated with
respect to the FASER simulation results~\cite{FASER_cross}.
Such factors, however, propagated and resulted in about four times
higher $\nu_\mu$ yield in the light of the FASER number in simulation,
which also reflected onto the observed difference between the FASER
data and our $\nu_\mu$ simulation result. This level of agreement,
however at this point in time, does not bring down too much the
novelty of our estimations shown in
figures~\ref{FIG:N_cc},~\ref{FIG:N_W},~\ref{FIG:N_cc_1100},
and~\ref{FIG:N_W_1100}.

Note that we observed that the event generation times in
\textsc{Pythia8} are about 67 and 47 times faster than those in
\texttt{EPOS.LHC-R} at the center-of-mass energies at 13.6 and
100~TeV, respectively, in the validations.

\section{Summary}
Using a hypothetical detector configuration similar to  
FASER$\nu$, positioned either 0.5~km or 2~km downstream of the $pp$
interaction point along the beamline, and a tuned \textsc{Pythia8}
simulation, we estimated  forward neutrino production and event rates
at the proposed FCC-hh, operating at a $pp$ collision energy of
100~TeV and an integrated luminosity of 1~ab$^{-1}$.
Our findings indicate that charged current neutrino interaction
rates can be expected for neutrino energies reaching up to
approximately 50~TeV. Moreover, the FCC-hh may offer an 
opportunity to observe direct $W^-$ and $W^+$ boson production via
neutrino-nucleus interactions, a process that remains to be
experimentally quantified.

The approach outlined herein can be straightforwardly extended to
other detectors with different target masses or alternative
placements, through appropriate adjustments to the target geometry as
demonstrated in figures~\ref{FIG:N_cc_1100}
and~\ref{FIG:N_W_1100}. Such extensions indicate that even with the
rather conservative expectation results from \texttt{EPOS.LHC-R}, the
charged current neutrino interaction rates can be expected for
neutrino energies reaching up to approximately 50~TeV and provide a
chance to observe direct $W^-$ and $W^+$ boson production via
neutrino-nucleus interactions for the first time.

Therefore, our estimations from two different event generators
indicate that the FCC-hh will be an excellent place to test the SM for
BSM physics searches through neutrino-nucleus interactions employing
the entire FASER$\nu$-like detector.

\section*{Acknowledgments}
This work was supported by a Korea University Grant, the National
Research Foundation of Korea (NRF) grants funded by the Korea
government (MSIT) (RS-2022-NR068913, RS-2025-00556247, and
RS-2022-00143178), and the Korea Basic Science Institute (National
research Facilities and Equipment Center) grant funded by the Korea
government (MSIT) (NFEC-2019R1A6C1010027).

\vspace{0.2cm}
\noindent

\let\doi\relax

\end{document}